\documentclass[12pt,twoside]{article}
\usepackage{psfig}
\newcommand{\VolumeHeader}{}
\newcommand{\VolumeSerial}{LNS}

\newcommand{\ActivityName}{ {\normalsize {\it 
Gravitational Waves: A Challenge to Theoretical Astrophysics.}
}}
\newcommand{\ActivityDate}{ {\normalsize {\it
Trieste, 5-9 June 2000}
}}

\renewcommand\({\left(}
\renewcommand\){\right)}
\renewcommand\[{\left[}
\renewcommand\]{\right]}

\def\lsim{\raise 0.4ex\hbox{$<$}\kern -0.8em\lower 0.62
ex\hbox{$\sim$}}
\def\gsim{\raise 0.4ex\hbox{$>$}\kern -0.7em\lower 0.62
ex\hbox{$\sim$}}
\newcommand{\mpl}{M_{\rm Pl}}

\newcommand{\ogw}{\Omega_{\rm gw}}
\newcommand{\hogw}{h_0^2\Omega_{\rm gw}}
\newcommand{\hc}{h_c(f)}
\newcommand{\hn}{h_n(f)}
\newcommand{\hf}{\tilde{h}_f}

\newcommand\eq[1]{eq.~(\ref{#1})}
\newcommand\eqs[2]{eqs.~(\ref{#1}) and (\ref{#2})}

\newcommand\ee{\end{equation}}
\newcommand\be{\begin{equation}}
\newcommand\eea{\end{eqnarray}}
\newcommand\bea{\begin{eqnarray}}
\newcommand\sub[1]{_{\rm #1}}

\newcommand{\LectureHeader}{Stochastic backgrounds of GWs}

\begin{document}
\pagestyle{myheadings}
\markboth{\LectureHeader}{\VolumeHeader}
\markright{\VolumeHeader}


\begin{titlepage}


\title{Stochastic backgrounds of gravitational waves}

\author{M. Maggiore 
\\[1cm]
{\normalsize
{\it INFN and Dipartimento di Fisica, Pisa, Italy.}}
\\[10cm]
{\normalsize {\it Lecture given at the: }}
\\
\ActivityName 
\\
\ActivityDate 
\\[1cm]
{\small \VolumeSerial} 
}
\date{}
\maketitle
\thispagestyle{empty}
\end{titlepage}

\baselineskip=14pt
\newpage
\thispagestyle{empty}


\begin{abstract}

We review the motivations for the search of 
stochastic backgrounds of gravitational waves 
and we compare the  experimental sensitivities that can be
reached in the near future with the existing bounds and with the
theoretical predictions.

\end{abstract}

\vspace{6cm}

{\it Keywords:} Gravitational Waves

{\it PACS numbers:}
04.30.-w; 04.30.Db; 04.80.Nn; 98.80.Cq


\newpage
\thispagestyle{empty}
\tableofcontents

\newpage
\setcounter{page}{1}

\section{Motivations}

A possible target of gravitational wave (GW) experiments is given by
stochastic backgrounds of cosmological origin. In a sense, these are
the gravitational analog of the 2.7~K microwave photon background and, 
apart from their obvious intrinsic interest, they would carry extraordinary
informations on the state of the very early universe and on physics
at correspondingly high energies. To understand this point, one
should have in mind the following basic physical principle:
\begin{itemize}
\item {\it a background of relic particles gives a snapshot of the state of
the universe at the time when these particles decoupled from the
primordial plasma}. 
\end{itemize}
The smaller is the cross section of a particle,
the earlier it decouples. Therefore particles with only gravitational
interactions, like gravitons and possibly other fields predicted by
string theory, decouple much earlier than particles which have  also
electroweak or strong interactions. The condition for
decoupling is that the interaction rate of the process that
mantains equilibrium, $\Gamma$, becomes smaller than the
characteristic time scale, which is given by the Hubble parameter $H$,
\be
\Gamma \ll H \hspace{10mm}\Rightarrow {\rm decoupled}
\ee
(we set  $\hbar =c=1$).
A simple back-of-the-envelope computation shows
that, for gravitons,
\be
\left(\frac{\Gamma}{H}\right)=
\left(\frac{T}{\mpl}\right)^3\, ,
\ee
so that gravitons are decoupled
below the Planck scale $\mpl \sim 10^{19}$ GeV, {\it i.e.},
already $10^{-44}$ sec after the big-bang. This means that
a  background of GWs produced in
the very early universe encodes still today, in its frequency spectrum, 
all the informations about the conditions in which it was created.

For comparison, the photons
that we observe in the CMBR decoupled when the temperature was of
order $T\simeq 0.2$ eV, or $3\cdot 10^5$ yr after the big bang. This
difference in scales  simply reflects the difference in the strength
of the gravitational and electromagnetic interactions. Therefore,
the photons of the CMBR give us a snapshot of the state
of the universe at $t\sim 3\cdot 10^5$ yr. Of course, 
from this snapshot we can also understand many things about much earlier
epochs. For instance, 
the density fluctuations present at this epoch have been produced
much earlier, and the recent Boomerang data indicate that they
are compatible with the prediction from
inflation. Therefore, from the photon microwave background
we can extract informations on the state 
of the Universe at
much earlier times than the epoch of photon decoupling. In this case,
conceptually the situation is similar with trying to understand
the aspect that a person had as a child from a picture taken when he
was much older. Certainly many features can be inferred from such a
picture. Gravitational waves, however, provide directly a picture of
the child, and therefore give us really unique informations. 

It is clear that the reason why GWs are potentially so interesting is due to
their very small cross section, and the very same reason is at the basis of
the difficulty of the detection.
In the next sections we will review the experimental aspects of
the search for stochastic backgrounds and the theoretical
expectations. Details and more complete references can be found
in ref.~\cite{MM}.

\section{Characterization of stochastic backgrounds of GWs}\label{sec2}

The intensity of  a stochastic background of GWs
can be characterized by the dimensionless quantity
\be
\ogw (f)=\frac{1}{\rho_c}\,\frac{d\rho_{\rm gw}}{d\log f}\, ,
\ee
where $\rho_{\rm gw}$ is the energy density of the stochastic
background of gravitational waves, $f$ is the frequency ($\omega =2\pi
f$) and $\rho_c$ is the present value of the
critical energy density for closing the universe. In terms of the present
value of the Hubble constant $H_0$ and Newton's constant $G_N$, 
the critical density is given by
$\rho_c =3H_0^2/(8\pi G_N)$.
The  value of $H_0$ is usually written as $H_0=h_0\times 100 $
km/(sec--Mpc), where $h_0$ parametrizes the existing experimental
uncertainty. However, 
it is not very convenient to normalize $\rho_{\rm gw}$
to a quantity, $\rho_c$,
which is uncertain: this uncertainty would appear
in all the subsequent formulas, although it has nothing to do with the 
uncertainties on the GW background itself. 
Therefore, we
rather characterize the stochastic GW background with
the quantity $\hogw (f)$, which is independent of $h_0$. All theoretical
computations of a relic GW spectrum are actually  computations of 
$d\rho_{\rm gw}/d\log f$ and  are  independent of the
uncertainty on $H_0$. Therefore the result of these computations is
 expressed in terms of $\hogw$, rather than of
$\ogw$.

To understand the effect of the stochastic background on a detector,
we need however to think in terms of amplitudes of GWs.
A stochastic GW at a given point $\vec{x}=0$ can be expanded, in the
transverse traceless gauge, as
\be\label{hab}
h_{ab}(t)=\sum_{A=+,\times}\int_{-\infty}^{\infty}df\int d\hat{\Omega}
\,\tilde{h}_A(f,\hat{\Omega})\, e^{-2\pi ift\, }e_{ab}^A(\hat{\Omega})\, ,
\ee
where $\tilde{h}_A(-f,\hat{\Omega})=\tilde{h}_A^*(f,\hat{\Omega})$. 
$\hat{\Omega}$ is a unit vector representing the direction of
propagation of the wave and $d\hat{\Omega}=d\cos\theta d\phi$.
The polarization tensors can be written as
$
e_{ab}^+(\hat{\Omega})=\hat{m}_a\hat{m}_b-\hat{n}_a\hat{n}_b\, ,
e_{ab}^{\times}(\hat{\Omega})=\hat{m}_a\hat{n}_b+\hat{n}_a\hat{m}_b\, ,
$
with $\hat{m},\hat{n}$ unit vectors ortogonal to 
$\hat{\Omega}$ and to each other. With these definitions,
$e^A_{ab}(\hat{\Omega})e^{A',ab}(\hat{\Omega})=2\delta^{AA'}$. 
For a
stochastic background,  assumed to be isotropic, unpolarized and
stationary, we can define the {\em spectral density}
 $S_h(f)$ from
the ensemble average of the Fourier amplitudes,
\be\label{ave}
 \langle \tilde{h}_A^*(f,\hat{\Omega})
\tilde{h}_{A'}(f',\hat{\Omega}')\rangle =
\delta (f-f')\frac{1}{4\pi}
\delta^2(\hat{\Omega},\hat{\Omega}')\delta_{AA'}
\frac{1}{2}S_h(f)\, ,
\ee
where $\delta^2(\hat{\Omega},\hat{\Omega}')=\delta (\phi -\phi ')
\delta (\cos\theta -\cos\theta ')$. $S_h(f)$
 has dimensions Hz$^{-1}$ and  satisfies $S_h(f)=S_h(-f)$.  The choice
of normalization is such that
\be\label{norm}
\int d\hat{\Omega}d\hat{\Omega}'\,
\langle \tilde{h}_A^*(f,\hat{\Omega})
\tilde{h}_{A'}(f',\hat{\Omega}')\rangle =
\delta (f-f')\delta_{AA'}
\frac{1}{2}S_h(f)\, .
\ee
$\ogw$ and $S_h$ are related by
\be\label{ooo}
\ogw (f)=\frac{4\pi^2}{3H_0^2} f^3S_h(f)\, .
\ee
In general, theoretical predictions are 
expressed more naturally
in terms of $\hogw (f)$, while the equations involving the
signal-to-noise ratio and other issues related to the detection are
much more transparent when written in terms of
$S_h(f)$. Eq.~(\ref{ooo}) is the basic formula for moving
between the two descriptions. 

The characteristic amplitude $h_c(f)$ is instead defined from 
\be\label{hc2}
\langle h_{ab}(t)h^{ab}(t)\rangle =
2 \int_{f=0}^{f=\infty}d(\log f)\,\, h_c^2(f)\, .
\ee
$h_c(f)$ is dimensionless, and represents a characteristic
value of the amplitude, per unit logarithmic interval of frequency.
The factor of two on the right-hand side of eq.~(\ref{hc2}) is
part of the  definition, and is motivated by the fact that the
left-hand side is made up of two  contributions,
given by
$\langle \tilde{h}_+^*\tilde{h}_{+}\rangle$ and
$\langle \tilde{h}_{\times}^*\tilde{h}_{\times}\rangle$. In a
unpolarized background these contributions are equal, while 
the mixed term $\langle \tilde{h}_+^*\tilde{h}_{\times}\rangle$
vanishes, \eq{ave}.
The relation betheen $h_c$ and $S_h$ is
\be\label{hc3}
h_c^2(f)= 2fS_h(f)\, .
\ee
Actually, $\hc$ is not yet the most useful dimensionless quantity to
use for the comparison with experiments. In fact, any experiment
involves some form of binning over the frequency. In a total
observation time $T$, the resolution in frequency is $\Delta f=1/T$,
so one does not observe $\hogw (f)$ but rather 
\be
\int_f^{f+\Delta f}d(\log f)\,\, \hogw (f)\simeq \frac{\Delta f}{f}\hogw
(f)\, ,
\ee
and, since $\hogw (f)\sim f^3S_h(f)\sim f^2 h_c^2(f)$, 
it is convenient to define
\be\label{rho6}
h_c(f,\Delta f)=\hc \(\frac{\Delta f}{f}\)^{1/2}\, .
\ee
Using $1/(1\, {\rm yr})\simeq 3.17\times 10^{-8}$ Hz
 as a reference value for $\Delta f$, and
$10^{-8}$ as a reference value for $\hogw$, one finds
\be\label{hrms2}
h_c(f,\Delta f)=7.111\times 10^{-22}
\left(\frac{{\rm 1\, mHz}}{f}\right)^{3/2} 
\(\frac{\hogw (f)}{10^{-8}}\)^{1/2}
\(\frac{\Delta f}{3.17\times 10^{-8}\,{\rm Hz}}\)^{1/2}\, .
\ee

\section{Characterization of the detectors}

The response of a detector is characterized  by two important
quantities: the {\em strain sensitivity} $\hf$, 
which gives a measure of the
noise in the detector, and the pattern functions $F^A(\theta ,\phi )$,
which reflect the geometry of the detector.

\subsection{Strain sensitivity}

The total output of the detector $S(t)$ is in general 
of the form
\be
S(t)=s(t)+n(t)
\ee
where $n(t)$ is the noise and $s(t)$ is the contribution to the output
due to the gravitational waves. Both are taken to be dimensionless
quantities. If the noise is gaussian,
the ensemble average of the Fourier components  of the noise,
$\tilde{n}(f)$, satisfies
\be\label{Sn1}
\langle \tilde{n}^*(f)\tilde{n}(f')\rangle =
\delta(f-f')\frac{1}{2}S_n(f)\, 
\ee
(actually, non-gaussian noises are  potentialy very dangerous in GW
experiments, and must be carefully minimized or modelled).
The above equation defines the
function $S_n(f)$, with $S_n(-f)=S_n(f)$ and dimensions Hz$^{-1}$. 

The factor $1/2$ is  conventionally inserted in the definition
so that the total noise power is obtained integrating $S_n(f)$ over the
physical range $0\leq f<\infty$, rather than from $-\infty$ to
$\infty$,
\be\label{n2}
\langle n^2(t)\rangle = \int_0^{\infty}df\, S_n(f)\, .
\ee
The function $S_n$ is known as the spectral noise
density. This quantity is quadratic in the noise; it is therefore
convenient to take its square root and define 
the {\em strain sensitivity} $\tilde{h}_f$
\be
\tilde{h}_f\equiv\sqrt{S_n(f)}\, , 
\ee
where now $f>0$;  $\tilde{h}_f$ 
 is linear in the noise and has dimensions
Hz$^{-1/2}$.

\subsection{Pattern functions}

GW experiments are designed so that their scalar output $s(t)$ is
linear in the GW signal: if
$h_{ab}(t)$ is the metric perturbation in the transverse-traceless gauge,
\be
s(t)=D^{ab}h_{ab}(t)\, .
\ee
$D^{ab}$ is known as the {\em detector tensor}.
Using \eq{hab}, we write
\be\label{27}
s(t)=\sum_{A=+,\times}\int_{-\infty}^{\infty}df\int d\hat{\Omega}
\,\tilde{h}_A(f,\hat{\Omega})\, e^{-2\pi ift}\,
D^{ab} e_{ab}^A(\hat{\Omega})\, .
\ee
It is  convenient to define the {\em detector pattern
functions} $F_{A}(\hat{\Omega} )$,
\be\label{defF}
F_{A}(\hat{\Omega} )=D^{ab} e_{ab}^A(\hat{\Omega})\, ,
\ee
so that
\be\label{28}
s(t)=\sum_{A=+,\times}\int_{-\infty}^{\infty}df\int d\hat{\Omega}
\,\tilde{h}_A(f,\hat{\Omega})F_A(\hat{\Omega})e^{-2\pi ift}\, ,
\ee
and the Fourier transform of the signal, $\tilde{s}(f)$, is
\be
\tilde{s}(f)=\sum_{A=+,\times}\int d\hat{\Omega}
\,\tilde{h}_A(f,\hat{\Omega})F_A(\hat{\Omega})\, .
\ee
Explicit expressions for the pattern functions of various detectors
can be found in ref.~\cite{MM}.

\subsection{Single detectors}

For a stochastic background the average of $s(t)$ vanishes and, if we
have only one detector, the best we can do is to
consider the average of $s^2(t)$. Using \eqs{ave}{28},
\be\label{FSh}
\langle s^2(t)\rangle =
F\, \int_{0}^{\infty}df\, 
S_h(f)\, ,
\ee
where
\be\label{FF}
F\equiv
\int\frac{d\hat{\Omega}}{4\pi}\sum_{A=+,\times} F^A(\hat{\Omega} )
F^A(\hat{\Omega} )\, 
\ee
is a factor that gives a measure of the angular efficiency of the
detector. For interferometers $F=2/5$, while for cilindrical bars
$F=8/15$.

In a single detector a
stochastic background will  manifest itself as an eccess noise.
Comparing  eqs.~(\ref{n2}) and eq.~(\ref{FSh}), we see that 
the signal-to-noise ratio at frequency $f$ is
\be\label{SNR1}
{\rm SNR}=\[ \frac{F S_h(f)}{ S_n(f)}\]^{1/2}\, . 
\ee
We  use the convention that the SNR refers to the 
GW amplitude, rather than to the GW energy. Since 
$\hogw (f)\sim h_c^2(f)$
the SNR for the amplitude is the square root of the SNR for the
energy, hence the square root in \eq{SNR1}.

Using eq.~(\ref{ooo}) 
and $S_n(f)=\tilde{h}_f^2$, we can express this result in terms of
the minimum detectable value of $\hogw$, at a given level
of SNR, as
\be\label{single}
\hogw^{\rm min}(f)\simeq 10^{-2}\,\frac{({\rm SNR})^2}{F}  
\left(\frac{f}{\rm 100 Hz}\right)^3
\left(\frac{\tilde{h}_f}{10^{-22}{\rm Hz}^{-1/2}}
\right)^2\, .
\ee

\subsection{Correlated detectors}

A much better sensitivity can be obtained correlating two (or more) 
detectors.
In this case we  write the output $S_i(t)$ of the $i-$th detector 
as $S_i(t)=s_i(t)+n_i(t)$,
where  $i=1,2$ labels the detector, and
we consider the situation 
in which the GW  signal $s_i$ is much smaller than the noise $n_i$.
We  correlate the two outputs defining
\be\label{S}
S_{12}=\int_{-T/2}^{T/2}dt\int_{-T/2}^{T/2}dt'\,
S_1(t)S_2(t')Q(t-t')\, ,
\ee
where $T$ is the total integration time (e.g. one year) and $Q$ is a
real filter function. The simplest choice would be $Q(t-t')=\delta
(t-t')$. However, if one knows the form of the signal that one is
looking for, {\it i.e.} $S_h(f)$,
it is possible to optimize the form of the filter
function. It turns out that the optimal filter is given 
by~\cite{Mic,Chr,Fla,All,VCCO}
\be\label{optimal}
\tilde{Q}(f)=c \frac{\Gamma (f)S_h(f)}{S_n^{(1)}(f)S_n^{(2)}(f)}
\ee
with $c$ an arbitrary normalization constant; $S_h(f)$ is the spectral
density of the signal and $S_n^{(i)}(f)$ the noise
spectral densities of the two detectors. Since in general $S_h(f)$ is
not known {\it a priori}, one should perform the data analysis 
considering  a set of possible filters. For many cosmological
backgrounds, a simple set of power-like filters $S_h(f)\sim
f^{\alpha}$ should be adequate. Actually, many of the most interesting
cosmological backgrounds are expected to be approximately flat within
the window of existing or planned experiments.

The function $\Gamma (f)$ in \eq{optimal} is the (unnormalized)
overlap reduction function; it is defined as
\be\label{Gamma}
\Gamma (f) \equiv \int \frac{d\hat{\Omega}}{4\pi}\,
\[ \sum_A F^A_1(\hat{\Omega} )F^A_2(\hat{\Omega} )\]
\exp\left\{ 2\pi if\hat{\Omega}\cdot\frac{\Delta\vec{x}}{c}\right\}\, ,
\ee
where $\Delta \vec{x}$ is the separation between the two detectors.
It gives a measure of how well the two detectors are correlated; for
instance, if the
distance between them is much bigger than the typical wavelength of
the stochastic GWs, the two detectors do not really see the same
GW backgrounds, and noting is gained  correlating them. 
It also depends on the relative
orientation between them.

It is conventional to introduce the
(normalized) {\em overlap reduction function} $\gamma(f)$~\cite{Chr,Fla} 
\be\label{gammaGamma}
\gamma(f)=\frac{\Gamma(f)}{F_{12}}\, ,
\ee
where
\be\label{FF12}
F_{12}\equiv
\int\frac{d\hat{\Omega}}{4\pi}\sum_A F^A_1(\hat{\Omega} )
F^A_2(\hat{\Omega} )|_{\rm aligned}\, .
\ee
Here the subscript means that we must compute $F_{12}$ taking the two
detectors to be perfectly aligned, rather than with their actual
orientation. If the two
detectors are of the same type, e.g. two interferometers or two
cylindrical bars,  $F_{12}$ is the same as the constant
$F$ defined in eq.~(\ref{FF}).
The use of $\Gamma (f)$ is more convenient when we want
to write equations that hold independently of what detectors
(interferometers, bars, or spheres) are
used in the correlation (furthermore, in the case of the correlation
between an interferometer and the scalar mode of a sphere
$F_{12}=0$, so this normalization is impossible; then,  one just uses 
$\Gamma (f)$, which is the quantity that enters directly in the SNR).

The SNR for two correlated detectors, with
optimal filtering, is
\be\label{SNR2}
{\rm SNR}=\left[ 2 T\int_0^{\infty}df\, 
 \Gamma^2(f)
\frac{S_h^2(f)}{S_n^2(f)}
\right]^{1/4}\, .
\ee
Comparing with \eq{SNR1} we see that now the SNR is given by an
integral over all frequencies, rather then by a comparison of $S_h(f)$
with $S_n(f)$ at fixed $f$. We denote by
$\Delta f$  the frequency range where both detectors are sensitive
and at the same time the overlap reduction
function $\Gamma (f)$ is not much smaller than one. Then, if
the integration time $T$ is such that
$T\Delta f\gg 1$, correlating two detectors is much better
than working with the  single detectors.

Finally, it can be useful to define a dimensionless
characteristic noise associated
to the correlated detectors. It turns out that the quantity that is
meaningful to compare directly to the characteristic
dimensionless  amplitude $\hc$ of the GW (defined in sect.~\ref{sec2})
is a characteristic noise $\hn$ defined by
\be\label{hn}
h_n(f)\equiv\frac{1}{(\frac{1}{2}T\Delta f)^{1/4}}\,
\left(\frac{fS_n(f)}{\Gamma (f)}\right)^{1/2}\, .
\ee
The characteristic 
noise $\hn$ is a useful but approximate concept, since the correct
expression of the SNR for correlated detectors
is obtained integrating over all frequencies, 
\eq{SNR2}, and not comparing $\hn $ and $\hc$ at fixed $f$.

\section{Experiments}

\subsection{Resonant bars}

There are five ultracryogenic resonant bars around the world, in
operation since a few years:

\begin{table}[htb]
\begin{center}
\begin{tabular}{||l|c|r||}\hline
detector & location          & taking data since\\ \hline\hline
NAUTILUS & Frascati, Rome    & 1993\\ \hline
EXPLORER & Cern (Rome group) & 1990\\ \hline
ALLEGRO  & Louisiana, USA    & 1991\\ \hline
AURIGA   & Padua, Italy      & 1997\\ \hline
NIOBE    & Perth, Australia  & 1993\\ \hline
\end{tabular}
\caption{The existing resonant bars.}
\end{center}
\end{table}

Compared to interferometers, they are narrow band detectors, since
they operate in a frequency band of the order of a few Hz, with resonant
frequency of the order of 900 Hz (it should be observed, however,
that with recent improvement in the electronics, EXPLORER has now a
useful band of the order of a few tens of Hz). Fig.~1 shows the
sensitivity curve of NAUTILUS. 

\begin{figure}[htb]
\centerline{\hbox{ \psfig{figure=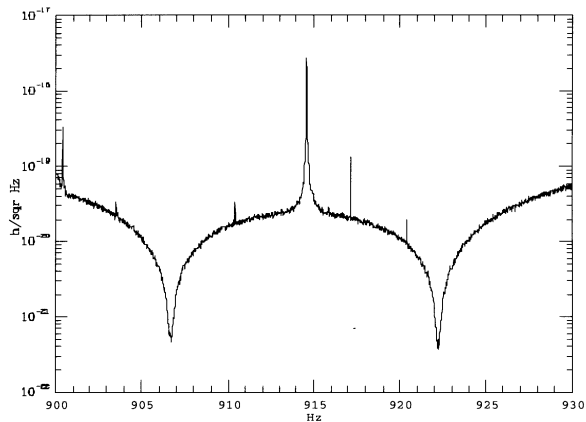,width=0.5\linewidth} }}
\caption{The sensitivity curve of NAUTILUS (courtesy of the NAUTILUS
  collaboration) }
\label{10_8_98}
\end{figure}

\begin{figure}[htb]
\centerline{\hbox{ \psfig{figure=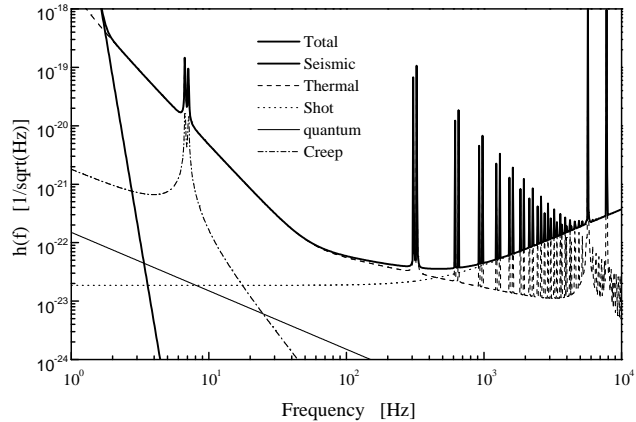,angle=270,width=0.5\linewidth} }}
\caption{The planned VIRGO sensitivity curve.}
\label{virsens}
\end{figure}

\subsection{Ground-based interferometers} 

The first generation of
large scale interferometers is presently under
construction. LIGO consists of two detectors, in Hanford, Washington,
and Livingston, Louisiana, with 4 km arms. VIRGO is under
construction  near Pisa, and has 3 km arms. Somewhat smaller are
GEO600, near
Hannover, with 600m arms and TAMA300 in Japan; these smaller
interferometers also aim at
developing advanced techniques needed for second-generation
interferometers. Fig.~\ref{virsens}
shows the planned VIRGO sensitivity curve.
Interferometers are wide-band detectors, and will cover the region
between a few Hertz up to approximately a few
kHz. Comparing the data in fig.~\ref{virsens} with \eq{single} we see that
used as single detectors, VIRGO and LIGO could measure
\be
\hogw\gsim 10^{-2}\, . 
\ee
As discussed in the previous section,
much more interesting values can be obtained
correlating two interferometers, an interferometer and a bar, or two
bars. The values for one year of integration time are given in  table~2.

\begin{table}[htb]
\begin{center}
\begin{tabular}{||l|l|c||}\hline
detector 1 & detector 2 & $\hogw$         \\ \hline\hline
LIGO-WA    & LIGO-LA    & $5\times 10^{-6}$\\ \hline
VIRGO      & LIGO-LA    & $4\times 10^{-6}$\\ \hline
VIRGO      & LIGO-WA    & $5\times 10^{-6}$\\ \hline
VIRGO      & GEO        & $5.6\times 10^{-6}$\\ \hline
VIRGO      & TAMA       & $1\times 10^{-4}$\\ \hline
VIRGO      & AURIGA     & $4\times 10^{-4}$\\ \hline
VIRGO      & NAUTILUS   & $7\times 10^{-4}$\\ \hline
AURIGA     & NAUTILUS   & $5\times 10^{-4}$\\ \hline
\end{tabular}
\caption{The sensitivity for various two-detectors correlations.}
\end{center}
\end{table}

\subsection{Advanced interferometers}

The interferometers presently
under construction are the first generation of large-scale
interferometers, and second generation interferometers, with much
better sensitivities, are under study. In particular,
the first data from the initial LIGO are expected by 2002, and the
improvements leading from the
initial LIGO to the advanced detector are expected to take place
around 2004-2006.

The overall improvement of LIGOII is expected to be, depending on the
frequency, one or two orders of magnitude in $\tilde{h}_f$. This is
quite impressive, since
two order of magnitudes in $\tilde{h}_f$ means four order of
magnitudes in $\hogw (f)$ and therefore interesting
sensitivity,  $\hogw$ of order  $10^{-5}$,
even for a single detector, without correlations. 

Correlating two advanced detectors one could reach extremely
interesting values:
the sensitivity of the correlation between two advanced LIGO is
estimated to be~\cite{All}
\be
\hogw\simeq 5\times 10^{-11}\, .
\ee

\subsection{The space interferometer LISA} 

This project,
approved by ESA and NASA, 
is an interferometer which will be sent into space around 2010.
Going into space, one is not limited
anymore by seismic and gravity-gradient noises; 
LISA could then explore the very low frequency
domain, $10^{-4}$~Hz $<f<1$~Hz. At the same time, there is also the
possibility of a very long path length (the mirrors will be freely
floating into the spacecrafts at distances of $5\times 10^{6}$ km from
each other!), 
so that the requirements on
the position measurement noise can be relaxed. The goal is to 
reach a strain sensitivity
$\tilde{h}_f=4\times  10^{-21} \,\, {\rm Hz}^{-1/2}$
at $f=1$ mHz. At this level, one expects first of all signals from
galactic binary
sources, extra-galactic supermassive black holes binaries and
super-massive black hole formation.
Concerning the stochastic background, going to low frequencies
provides a terrific advantage:  eq.~(\ref{ooo}) tells that
\be\label{ter}
\ogw (f)\sim f^3S_h(f)\, .
\ee
The ability to perform an interferometric measurement 
is encoded into the
spectral density of the noise, $S_n(f)$, and with a single detector
we can  measure $S_h(f)\sim S_n(f)$. Then, if we are able to reach
a good  value of $S_n$  at low frequency, \eq{ter} shows that we are
able to measure a very small value of $\ogw$, thanks to the factor
$f^3$. Going down by a factor $10^5$ in frequency, say from  10 Hz to 1
mHz, the factor $f^3$ provides an improvement by a factor
$10^{15}$! Even used as a single detector, LISA can therefore reach
very interesting sensitivities. Indeed, in terms of $\hogw$
the sensitivity of LISA corresponds to
\be\label{-12}
\hogw (f=1\,{\rm mHz})\simeq 1\times 10^{-12}\, .
\ee
We can now also understand why
it is instead very difficult to build a good detector
of stochastic GWs at high frequencies, $f\gg 1$~kHz. 
One can imagine to build
an apparatus that measures extremely small displacements, at high
frequencies, so that at some $f\gg 1$ kHz, the spectral density of the
noise 
$S_n(f)$ is very small. However, 
the quantity which is relevant for
the comparison with theoretical prediction, and on which strong
theoretical bounds exist at all frequencies, is $\hogw$, and
once  one translates the sensitivity in terms of 
$\hogw$, it will be very poor
because  the factor $f^3$ now works in the
wrong direction. 

\clearpage

\section{Bounds on $\hogw$}

\begin{figure}[htb]
\centerline{\hbox{
\psfig{figure=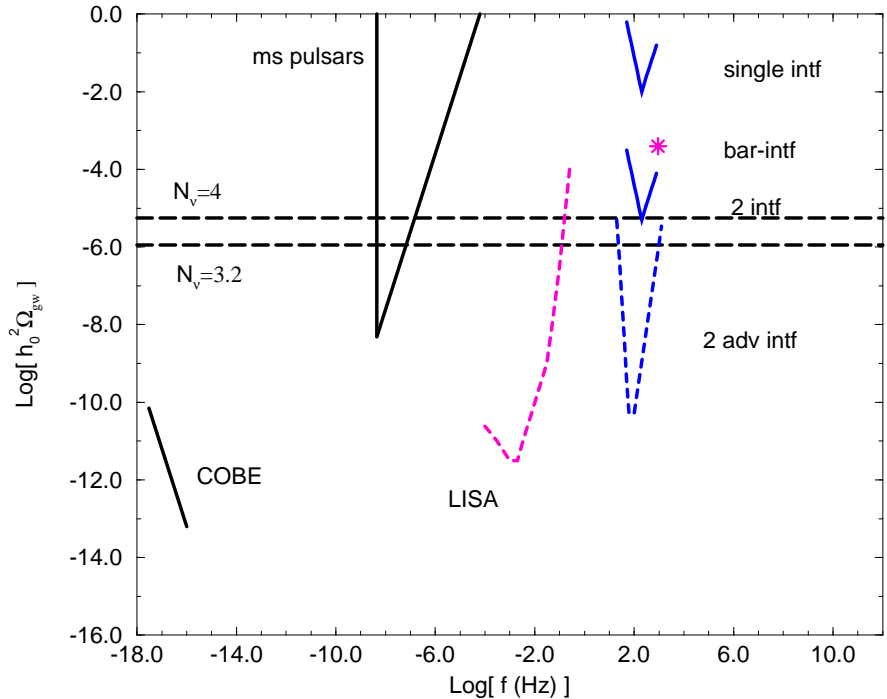,width=0.8\linewidth,angle=270 } }}
\caption{The bounds from nucleosynthesis (horizontal dashed lines,
for $N_{\nu}=4$ and for $N_{\nu}=3.2$), from COBE and from ms pulsars,
together with the sensitivity of a single ground based detector 
of the VIRGO or LIGO type (blue solid line), 
a bar-interferometer correlation (pink star), a
2-interferometer correlation (blue solid line; 
LIGO-LIGO, LIGO-VIRGO and VIRGO-GEO all
give very similar sensitivities), two advanced LIGO correlation
(blue dashed line ) and LISA (pink dashed line).}
\label{theo}
\end{figure}

In Fig.~\ref{theo} we show the most relevant bounds on  
stochastic GW backgrounds, together with the experimental
sensitivities discussed above. On the
horizontal axis we cover a huge range of frequencies. The lowest value,
$f=10^{-18}$ Hz, corresponds to a wavelength as large as the present Hubble
radius of the universe; the highest value shown, $f=10^{12}$ Hz, has instead
the following meaning: if we take a graviton produced during the
Planck era, with a typical energy of the order of the Planck or string
energy scale,
and we redshift it to the present time using the standard cosmological
model, we find that today it has a frequency of order $10^{11}$ or
$10^{12}$ Hz. This value therefore is the maximum possible
cutoff of  spectra of GWs produced in the very early universe. 
The maximum
cutoff for astrophysical processes is of course much lower, of order
10~kHz~\cite{Th,Sch}. So this huge frequency range encompasses all the
GWs that can be considered. 
Let us now discuss in turn the various bounds.

\subsection{Nucleosynthesis bound} 

The  outcome of nucleosynthesis depends on a  balance between
the particle production rates and the expansion rate of the
universe, measured by the Hubble parameter $H$. Einstein equation
gives $H^2\sim G_N\rho$, where $\rho$ is the total energy density,
including of course $\rho\sub{gw}$. 
Nucleosynthesis successfully predicts the primordial abundances of
deuterium, $^3$He, $^4$He and $^7$Li  
assuming that the only contributions to
$\rho$ come from the particles of the Standard Model, and no GW
contribution. 
Therefore, in order not to spoil the agreement, any further
contribution to $\rho$ at time of nucleosynthesis, including the
contribution of GWs, cannot exceed a maximum
value. The bound is usually written in terms of an effective number of
neutrino species $N_{\nu}$, and, applied to GWs, reads
\be\label{nsbound}
\int_{f=0}^{f=\infty}d(\log f)\,\, \hogw (f)\leq
5.6\times 10^{-6}  (N_{\nu}-3)\, .
\ee
The limit on $N_{\nu}$  is subject to various
systematic errors, which have to do mainly with the
issues of how much of the observed
$^4$He abundance is of primordial origin,
and of the nuclear processing of $^3$He in stars. Different bounds on
$N_{\nu}$ have therefore been proposed. The 
analysis of ref.~\cite{BNTT}
gives $N_{\nu}<3.2$ at 95\% c.l.;
this is shown in fig.~\ref{theo}, together with a more conservative
bound  $N_{\nu}<4$. Note that the nucleosynthesis 
bound is really a bound over
the total energy density, i.e. over 
the integral of $\hogw (f)$ over $d\log
f$. However, if the integral cannot exceed this value, also its
positive integrand cannot exceed it over a sizable range of
frequencies. The actual bound on $\hogw (f)$ depends on its frequency
dependence, and if for instance  $\hogw (f)$ is approximatly constant
between two frequencies $f\sub{min}$ and $f\sub{max}$, the bound on
$\hogw$ is stronger by a factor $\sim\log (f\sub{max}/ f\sub{min})$.
Of course, the nucleosynthesis bound applies only to GW produced
before nucleosynthesis, and is not relevant for GW of astrophysical
origin.

\subsection{Bounds from millisecond pulsars} 

Millisecond pulsar are an extremely impressive source of high
precision measurements~\cite{Lor}. For instance, the  observations of the
first msec pulsar discovered, B1937+21, after 9 yr of data, 
give a period of  
$1.557\, 806\, 468\, 819\, 794\, 5\pm
 0.000\, 000\, 000\, 000\, 000\, 4$ ms. As a consequence,
pulsars are also a natural  detector of
GWs, since a GW passing between us and the pulsar 
causes a fluctuation in the time of arrival of the pulse,
proportional to the GW amplitute $\hc$.
If the uncertainty in the time of
arrival of the pulse is $\epsilon$ and the total observation time is
$T$, this `detector' is sensitive to $\hc\sim \epsilon/T$, for
frequencies $f\sim 1/T$. The highest sensitivities can then
be reached for a continuous source, as a stochastic background,  after
one or more years of integration, and therefore for $f\sim
10^{-9}-10^{-8}$ Hz. Based on the data from PSR B1855+09, 
ref.~\cite{TD}
gives a limit, at $f\equiv f_*=4.4\times 10^{-9}$ Hz
(at 90\% c.l.),
\be
\hogw (f_*)<  4.8\times 10^{-9}\, .
\ee
Since the resolution on $\hc$ is proportional to $1/T$ and $\hogw\sim
h_c^2$, the bound for $f>f_*$ is 
\be
\hogw (f)<  4.8\times 10^{-9}\, \(\frac{f}{f_*}\)^2\, ,
\ee
and therefore it is quite significant (better than the nucleosynthesis 
bound)
even for $f\sim 10^2 f_*$. For $f<f_*$, instead, the pulsar provides
no limit at all.  With the observation time 
the bound will improve steadily and will move toward lower and lower
frequencies. It is given by the wedge-shaped curve in
fig.~\ref{theo}.

\subsection{Bound from COBE} 

Another important constraint comes from the COBE measurement of the
fluctuation of the temperature of the cosmic microwave background
radiation (CMBR). 
The basic idea is that
a strong background of GWs at very long wavelengths
produces a stochastic redshift on the frequencies of the photons of
the 2.7K radiation, and therefore a fluctuation 
in their temperature (Sachs-Wolfe effect).
The analysis of refs.~\cite{AKo,All}, where
the effect of multipoles with $2\leq l\leq 30$ is included, 
gives a bound 
\be\label{cobe2}
\hogw (f)<
7\times 10^{-11}\(\frac{H_0}{f}\)^2\, ,
\hspace*{20mm}(3\times 10^{-18}\,{\rm Hz}<f<10^{-16}\,{\rm Hz})
\, .
\ee 
This bound  is  stronger at the upper edge of its range of
validity,  $f\sim 30 H_0\sim 10^{-16}$ Hz,
where it gives
\be\label{cobebound}
\hogw (f)< 7\times 10^{-14}\, ,\hspace*{20mm}(f \sim\, 10^{-16}\,\rm{Hz} )\, .
\ee

\section{Theoretical predictions}

\begin{figure}[htb]
\centerline{\hbox{
\psfig{figure=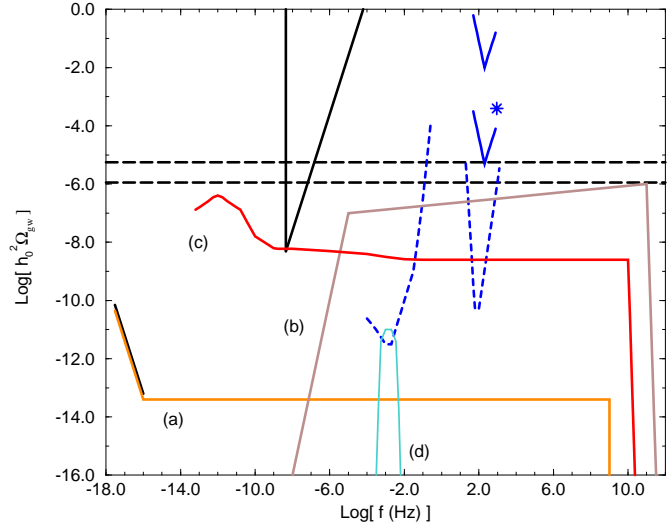,width=0.6\linewidth,angle=270 } }}
\caption{The backgrounds predicted, with optimistic choice of
parameters,  by (a) inflation,
(b) string cosmology, (c) cosmic strings, (d) a first order phase
transition at the electroweak scale, together with the bounds and
sensitivities of fig.~\ref{theo}.}
\label{predth}
\end{figure}

Many cosmological  production mechanisms have been proposed in 
recent years. Four examples are shown in fig.~\ref{predth}:

{\bf Inflation.} The amplification of vacuum fluctuations at the
transition between an inflationary phase and the radiation dominated
era produces GWs shown as the curve $(a)$ in  fig.~\ref{predth} and
is one of the oldest~\cite{Gri} and most studied examples.
The condition that the COBE
bound is not exceeded puts a  limit on the value of 
$\hogw$ at all frequencies, which is below the experimental
sensitivities, even for LISA.  
So, while it is one of the best studied examples,
it appears that it is not very  promising from the point of view of
detection. 

\vspace{3mm}

{\bf String cosmology.} In a cosmological model which follows
from the low energy action of string theory~\cite{Ven,GV} the
amplification of vacuum fluctuations can give a much stronger
signal~\cite{BGGV,peak}. The model has two free parameters that
reflect our ignorance of the large curvature phase.
The curve $(b)$ in fig.~\ref{predth} shows
the very interesting signal that could be obtained for some choice of
these parameters, while for other choices the value of $\hogw$ at
VIRGO/LIGO or at LISA frequencies becomes unobservably small.

\vspace{3mm}

{\bf Cosmic strings.} These are topological defects that can exist in
grand unified theories~\cite{VS}, and  vibrating, they produce a
large amount of GWs, shown in curve $(c)$ of fig.~\ref{predth}. 
Cosmic strings are characterized by a mass per unit length $\mu$, 
and the most stringent bound on $G_N\mu$ comes  from msec pulsars, 
and it is of order $10^{-6}$.

\vspace{3mm}

{\bf Phase transitions.} Another possible source of GWs is given by
phase transitions in the early universe. In particular, a phase
transition at the electroweak scale would give a signal just in the
LISA frequency window, while the QCD phase transition is expected to
give a signal peaked around $f=4\times 10^{-6}$ Hz. However, the
signal is sizable only if the phase transition is first order and,
unfortunately, in the Standard Model with the existing bounds on the
Higgs mass, there is not even a phase transition but rather a
smooth crossover, so that basically no GW is produced. However, in
supersymmetric extensions of the Standard Model, the transition can be
first order, and a stronger signal could be obtained. Depending on the
strength of the transition, one could even get a signal such as curve 
$(d)$ of fig.~\ref{predth}.  

\begin{figure}[htb]
\centerline{\hbox{
\psfig{figure=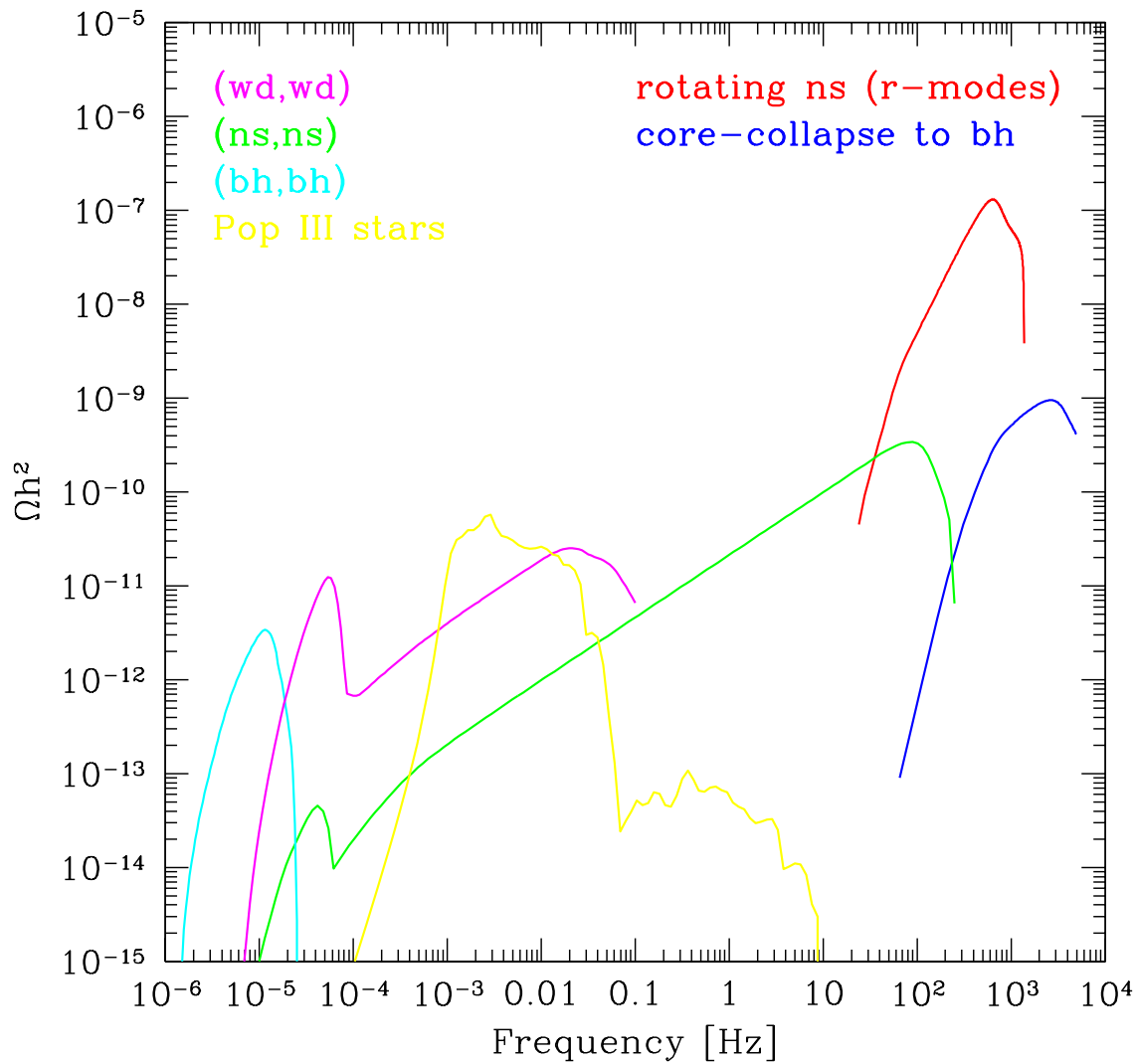,width=\linewidth,angle=0 } }}
\caption{Stochastic backgrounds of astrophysical origin. From left to
right, black hole-black hole binaries,
white dwarf-white dwarf binaries, neutron star-neeutron star binaries
(the curve extending up to $10^2$), a pregalactic
Population III stars, r-modes of neutron stars and 
supernova collapse to black holes
(figure kindly
provided by Raffaella Schneider).}
\label{astro}
\end{figure}

\vspace{3mm}

Finally, there are very interesting astrophysical backgrounds, coming
from a large number of unresolved sources. These are displayed in 
fig.~\ref{astro}. For a discussion, see \cite{FMS} and
the contribution of Raffaella Schneider to these proceedings.
Another important issue, especially for LISA, is also how to
discriminate cosmological from astrophysical backgrounds, see
eg.~\cite{UV}. 

The conclusion that emerges looking at fig.~\ref{theo}
is that in the next few years, with the first
generation of ground based interferometers, we will
have the possibility to explore five new
order of magnitude in energy densities, probing the content in GWs of
the universe down to $\hogw\sim 10^{-5}$. At this level, the
nucleosynthesis bound suggest that the possibility of detection are
quite marginal. It should not be forgotten, however, that
nucleosynthesis is a (beautiful) theory, with a lot of theoretical
input from nuclear reaction in stars, etc., and its prediction is by
no means a substitute for a measurement of GWs.
With the second generation of ground based interferometers and with
LISA, we will then penetrate quite deeply into a region which
experimentally is totally unexplored, and where a number of explicit
examples (although subject to large theoretical uncertainties) suggest
that a positive result can be found.

\clearpage


\end{document}